\documentclass[twocolumn,showpacs,amsmath,amssymb,pra]{revtex4}

\usepackage{graphicx}
\usepackage{dcolumn}
\usepackage{bm}
\usepackage[squaren, thinqspace]{SIunits}
\usepackage{nicefrac}

\usepackage[utf8]{inputenc}

\usepackage[plainpages=false, bookmarks=false, pdftoolbar=true, %
    pdfmenubar=true, pdffitwindow, %
    pdftitle={EIT Things}, pdfauthor={Atreju Tauschinsky}, %
    colorlinks=true, pdfhighlight=/O, pdfstartview=FitH, linkcolor=blue, %
    citecolor=red, urlcolor=blue]{hyperref}
\usepackage{amsmath}
\usepackage{epstopdf}

\begin{document}

\title{Spatially Resolved Excitation of Rydberg Atoms and Surface Effects on an Atom Chip}

\author{Atreju Tauschinsky}%
\email{Atreju.Tauschinsky@uva.nl}
\author{Rutger M. T. Thijssen}%
\author{S. Whitlock}%
\author{H. B. van Linden van den Heuvell}%
\author{R. J. C. Spreeuw}
\affiliation{Van der Waals-Zeeman Institute, University of Amsterdam, %
Valckenierstraat 65, 1018 XE Amsterdam, The Netherlands.}

\date{\today}   

\begin{abstract}
We demonstrate spatially resolved, coherent excitation of Rydberg atoms on an atom chip. Electromagnetically induced transparency (EIT) is used to investigate the properties of the Rydberg atoms near the gold coated chip surface. We measure distance dependent shifts ($\sim$\unit{10}{\mega\hertz}) of the Rydberg energy levels caused by a spatially inhomogeneous electric field. The measured field strength and distance dependence is in agreement with a simple model for the electric field produced by a localized patch of Rb adsorbates deposited on the chip surface during experiments. The EIT resonances remain narrow (\unit{<4}{\mega\hertz}) and the observed widths are independent of atom-surface distance down to $\sim$\unit{20}{\micro\meter}, indicating relatively long lifetime of the Rydberg states. Our results open the way to studies of dipolar physics, collective excitations, quantum metrology and quantum information processing involving interacting Rydberg excited atoms on atom chips.
\end{abstract}




\maketitle

\section{Introduction}

Ultracold atoms on atom chips are key to new atom-based technologies such as interferometers and precision sensors and provide access to fundamental aspects of many-body physics, atom-surface interactions, quantum metrology and quantum information science~\cite{Fortagh2007,Cronin2009}. So far experiments have dealt with atoms prepared in the electronic ground state, due to their intrinsic stability. Despite their weak interactions, ground-state atoms on atom chips have been used to sensitively probe the intrinsic thermal noise near surfaces~\cite{Jones2003, Harber2003, Zhang2005}, map magnetic and electric field distributions~\cite{Gunther2005, Wildermuth2005, Wildermuth2006, Whitlock2007, Obrecht2007, Sandoghdar1992}, and investigate the Casimir-Polder potential in the micrometer range~\cite{Lin2004, Harber2005, Bender2010a}. Comparatively, atoms excited to high-lying Rydberg states have extremely large transition dipole moments (scaling with $n^2$) resulting in long-range interactions and have large electric polarizabilities ($\propto n^7$) which can greatly enhance both atom-atom and atom-surface interactions.

Recent experiments with Rydberg atoms have largely been motivated by the excitation blockade mechanism~\cite{Tong2004, Heidemann2007}.
This made possible the experimental demonstration of a two-qubit quantum gate and entanglement between neutral atoms in separate microtraps~\cite{Urban2009, Gaetan2009}. Interactions can be further enhanced and controlled in the presence of modest electric fields via F{\"o}rster resonances~\cite{Foerster1948} over distances of tens of micrometers~\cite{Tauschinsky2008a, VanDitzhuijzen2009, VanDitzhuijzen2008a}. We aim to employ Rydberg blockade to optically control interactions between atomic ensembles prepared in separate magnetic microtraps on a magnetic lattice atom chip~\cite{Gerritsma2007, Whitlock2009}. Here, an unknown factor is the influence of the nearby metallic surface on the lifetime and coherence properties of the excited Rydberg atoms.

In this paper we report the coherent excitation of Rydberg atoms on an atom chip\cite{Mozley2005}. To probe the atom-surface potential and lifetime of the Rydberg states we employ excited-state electromagnetically induced transparency (EIT)~\cite{Mohapatra2007, Weatherill2008}. The position and width of the narrow transmission resonance reflect the energy and lifetime of the Rydberg state and provide a sensitive probe of the atom-surface interaction. This is related to a recent experiment investigating Rydberg excitation in a room temperature glass vapour cell at very small distances to the walls~\cite{Kuebler2010}. Here we investigate the effects of a metallic atom-chip surface with ultracold Rydberg atoms and with much higher energy resolution.

\section{Electromagnetically Induced Transparency}
\begin{figure}[htb]
    \includegraphics[width=1.00\columnwidth]{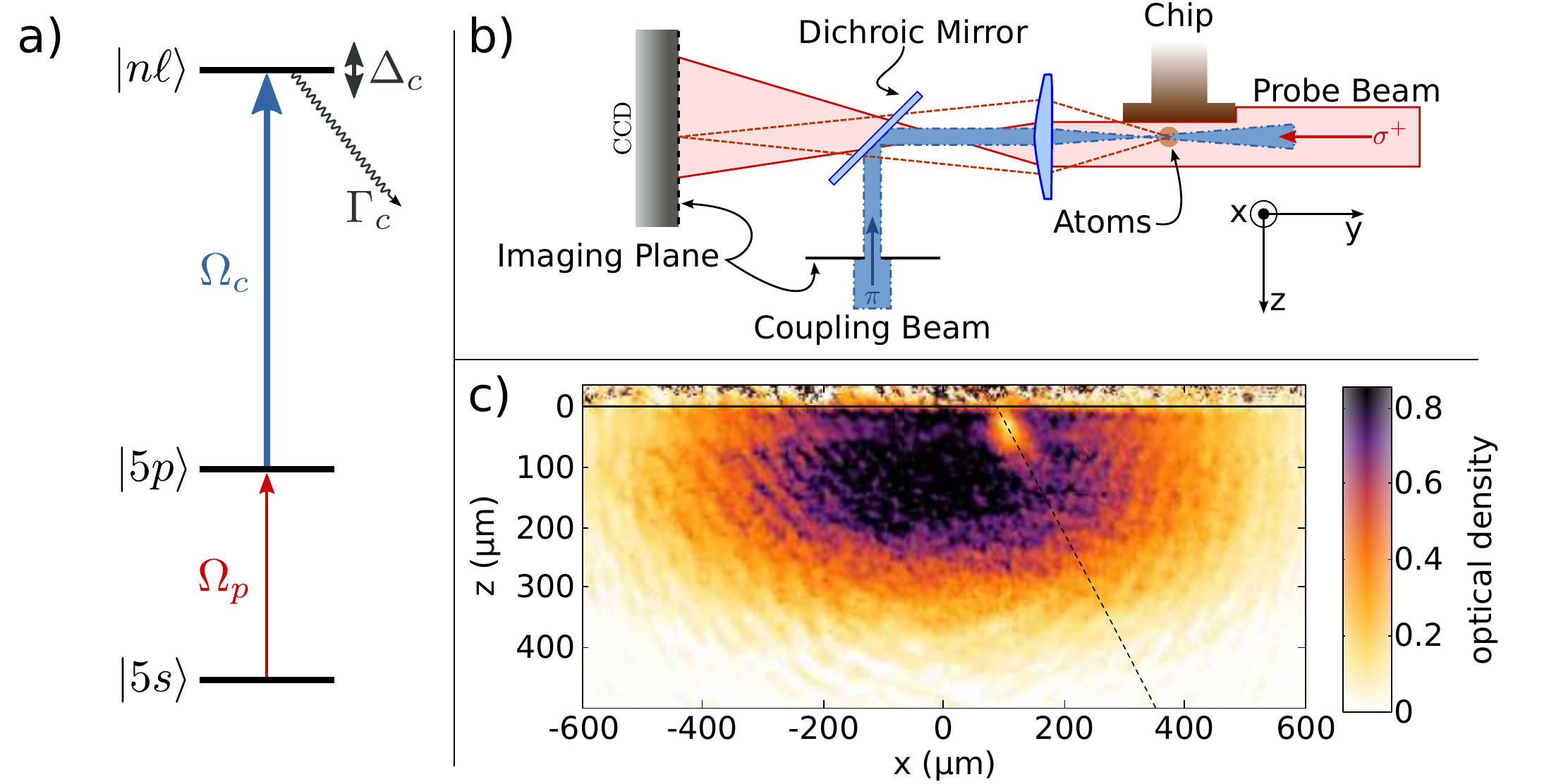}
    \caption{(Color online) (a) Level scheme of our system; shifts in the Rydberg state are denoted by $\Delta_c$, additional decay channels lead to a broadening of the Rydberg transition, i.e. an increase in $\Gamma_c$. (b) Schematic picture of the experimental Setup. Probe beam, coupling beam and the imaging system are depicted. (c) Typical absorption image taken on the $5s-5p$ transition. The chip is located at the top of the figure. The induced transparency due to the coupling laser is visible around $x=$~\unit{100}{\micro\meter}.}
    \label{fig:setup}
\end{figure}

Electromagnetically induced transparency is a coherent interference effect where the absorption on a resonant transition between two states is strongly modified by a coupling to a third state, creating a narrow transparency window. A common EIT configuration~\cite{Shepherd1996, Gea-Banacloche1995} is the ladder-type system investigated here. We probe the absorption on the $5s$-$5p$ transition of $^{87}$Rb, with the $5p$-state strongly coupled with a resonant laser to a highly excited $nd$ or $ns$ Rydberg state, as is depicted in figure \ref{fig:setup}a.

In this experiment the coupling laser is frequency stabilized directly to the $5p\rightarrow n\ell$-transition such that the transparency window is near the center of the resonance. In the limit of low probe intensity the susceptibility is given by
\begin{equation}
\chi(\Delta_p) \propto \frac{i \Gamma_p}{\Gamma_p + 2i\Delta_p + \frac{\Omega_c^2}{\Gamma_c + 2i\left(\Delta_p + \Delta_c\right)}},
\label{eqn:eitmodel}
\end{equation}
where the probe absorption is proportional to the imaginary part Im$(\chi)$~\cite{Gea-Banacloche1995}. In this equation $\Gamma_p$ and $\Gamma_c$ denote the decay rates of the probe- and coupling resonances respectively, $\Delta_p$ and $\Delta_c$ the detunings and $\Omega_c$ the Rabi frequency of the upper transition (Fig.~\ref{fig:setup}a).

If the Rydberg energy level shifts due to atom-atom or atom-surface interaction this leads to a detuning of the coupling laser from resonance which is visible as a shift of the transparency window in the absorption profile by a frequency $\Delta_c$. Furthermore, if the lifetime of the Rydberg state decreases, e.g. through induced decay to neighbouring states, this leads to a broadening of the transparency window, visible as an increase of $\Gamma_c$. Excited-state EIT is thus a sensitive probe to measure interactions in a Rydberg state. 

\section{Experimental Setup}

We trap approximately $7\times10^5$ $^{87}$Rb $|F=2, m_F=2\rangle$ atoms in a magnetic Ioffe-Pritchard type trap. The magnetic field required for this trap is produced by a Z-shape wire \unit{0.3}{\milli\meter} beneath the atom chip surface. The surface is coated with a \unit{100}{\nano\meter} gold layer and electrically isolated from the rest of the chip. The chip also features a permanent-magnet FePt layer that has been used to create arrays of microtraps~\cite{Gerritsma2007,Whitlock2009}, but this does not significantly influence the experiments discussed here. The atoms are cooled to temperatures of a few \micro\kelvin~by forced evaporative cooling. This setup is described in detail in~\cite{Whitlock2009}.

Prior to detection, the atoms are released from the trap and expand freely for \unit{2}{\milli\second} in a uniform magnetic field of \unit{1}{G} parallel to the probe beam. Typical peak atomic densities are $\unit{3\times10^9}{\centi\meter\rpcubed}$. After expansion the vertical extension of the cloud is \unit{100}{\micro\meter} and the cloud center is \unit{130(10)}{\micro\meter} from the surface (see figure \ref{fig:setup}c). The expanded cloud minimizes possible effects of atom-atom interactions and allows simultaneous probing of many atom-surface distances. 

We measure EIT spectra by simultaneously pulsing on a circularly polarized probe and a counterpropagating linearly-polarized (perpendicular to the chip surface) coupling laser beam for \unit{0.15}{\milli\second} and recording an absorption image of the probe with a CCD camera for a variable detuning $\Delta_p$. The resolution of our imaging system is $\sim$~\unit{7}{\micro\meter}. The probe beam is approximately uniform and much larger than the atom cloud, whereas the counterpropagating coupling laser beam is passed through an aperture and is imaged onto the atoms. The width of the coupling beam is $\sim$~\unit{60}{\micro\meter} and it extends $\sim$~\unit{150}{\micro\meter} from the chip surface; significantly smaller than the size of the atom cloud, i.e. only part of the cloud is exposed to the coupling beam and thus rendered transparent. This situation is depicted in figure \ref{fig:setup}b), and the spatial extent of the coupling beam is clearly visible as a light region (low optical density) centered around $x=$~\unit{100}{\micro\meter}, $z=$~\unit{50}{\micro\meter} in figure \ref{fig:setup}c.

The probe laser is frequency-stabilized directly to the $F = 2 \rightarrow F' = 3$ transition using Doppler-free polarization spectroscopy in a vapour cell. We estimate the linewidth of the laser around $\sim$~\unit{500}{\kilo\hertz} by beating it against an identical laser. To measure absorption spectra it is frequency shifted using a pair of double-pass AOMs which allow detunings between $\pm$~\unit{20}{\mega\hertz}. The coupling beam is produced by a frequency-doubled cw diode laser (Toptica TA-SHG). It is directly locked to the Rydberg state of interest by vapour cell EIT as described in~\cite{Raitzsch2009a}. 
This allows direct stable locking to both s- and d- Rydberg states in the range n = 19\ldots 70; the fine structure of the d-states is also well resolved in the spectroscopy. We estimate the combined (two-photon) linewidth of the Rydberg excitation lasers to be between 0.5 and \unit{2}{\mega\hertz}, depending on the Rydberg state used.

We extract spectra from a series of absorption images such as the one shown in figure \ref{fig:setup}c). By analyzing a series of such images taken for different detunings of the probe laser we construct an absorption spectrum as a function of detuning and distance to the surface. The coupling laser remains fixed in both position and frequency. For each detuning we take one image with the coupling laser present, and a reference image without the coupling laser. By analyzing each pixel of these images separately we obtain spatially resolved information about the transparency of the sample. In particular, this allows us to extract information for a large number of distances from the surface of the atom chip at once. The dashed line in figure \ref{fig:setup}c) denotes the peak EIT signal and is used to extract EIT spectra at different distances to the chip. A small angle between the beam profile and the surface normal helps to align the coupling beam parallel to the surface and minimize fringing.

\section{Rydberg Atoms on an Atom Chip}

\begin{figure}[htb]
    \includegraphics[width=1.00\columnwidth]{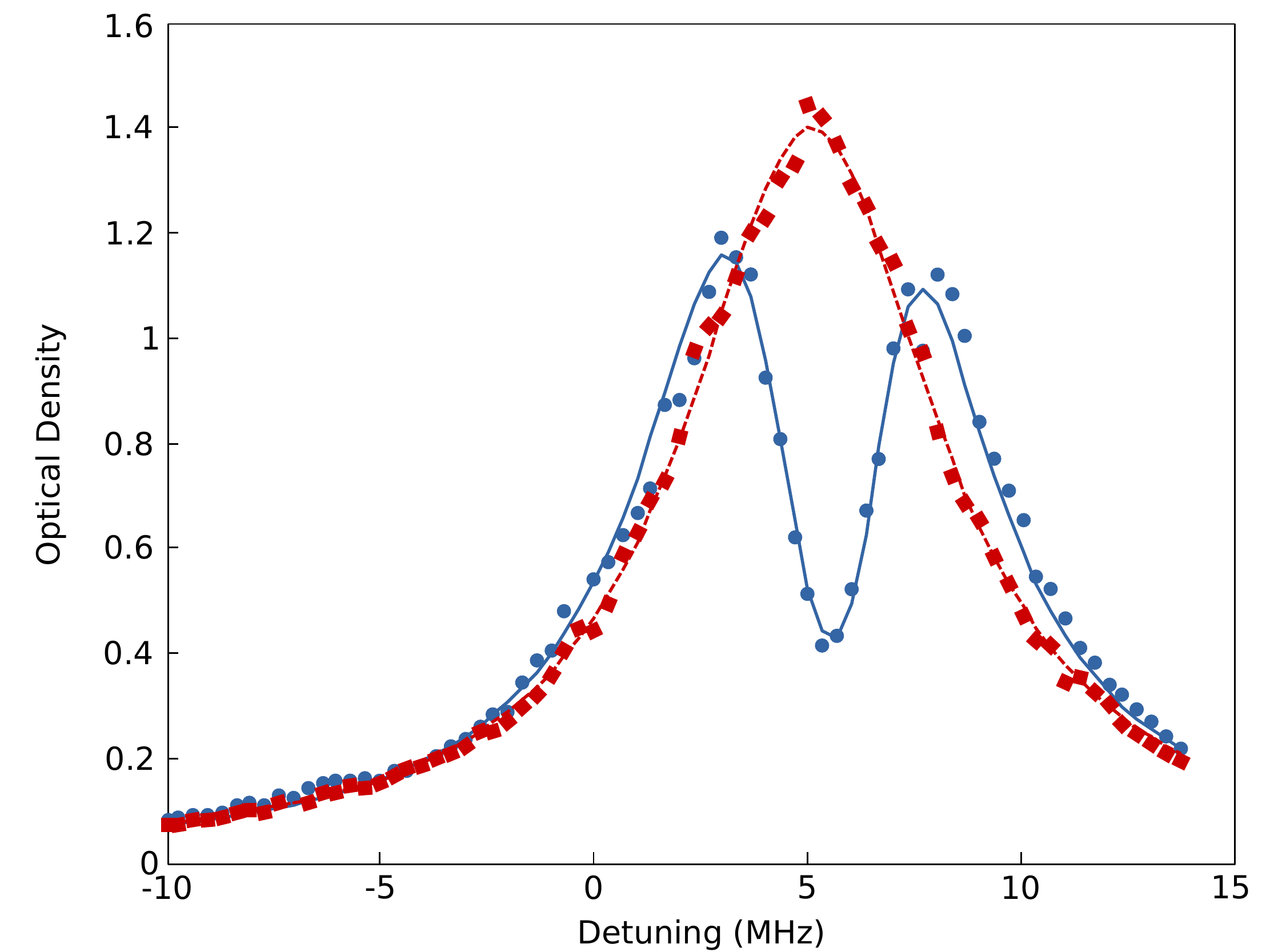}
    \caption{Optical spectra extracted from a series of absorption images for 80 probe detunings, with coupling to the 38d$_{5/2}$ Rydberg state. Square points indicate the measured optical density spectrum taken in the absence of the coupling laser, with a fit to Eq.~\eqref{eqn:eitmodel} with $\Omega_c=0$ (dashed line). Circles correspond to the EIT spectrum with the solid line a fit to the data using eq.~\ref{eqn:eitmodel}. Both spectra are obtained at a distance of $\sim$~\unit{200}{\micro\meter} from the surface of the chip.}
    \label{fig:spectrum}
\end{figure}

Figure~\ref{fig:spectrum} shows an absorption spectrum measured in the absence of the coupling beam (square points), along with a spectrum measured with the coupling beam present (circles), for a relatively large atom-surface distance of $z\sim$~\unit{200}{\micro\meter}. The former simply reflects the natural absorption lineshape of the $^{87}$Rb $5s\rightarrow5p$ transition and is used as a reference. The latter spectrum shows the EIT dip resulting from strong coupling to the $38d_{5/2}$ Rydberg state. The transparency window is close to the center of the $5s\rightarrow5p$ resonance. We can extract relevant parameters, in particular the linewidth and the detuning of the Rydberg state from these measurements by fitting the model of equation \eqref{eqn:eitmodel}. For the spectrum of figure~\ref{fig:spectrum} we find $\Delta_c =$~\unit{0.54(7)}{\mega\hertz} and $\Gamma_c =$~\unit{0.61(8)}{\mega\hertz}. The narrowest spectrum we have observed had a linewidth of \unit{0.4(3)}{\mega\hertz} for the state 23s$_{1/2}$ at a distance of \unit{115}{\micro\meter} from the chip, consistent with our estimates for the combined laser linewidth. The non-zero fitted value of $\Delta_c$ is due to a small offset in the locking of the probe- and coupling lasers, and constitutes a constant offset which is subtracted in later analysis. The measured EIT spectra demonstrate the coherent excitation of Rydberg atoms on an atom chip, in the distance range relevant to typical atom chip experiments. The width of the resonance is currently determined by the linewidth of the excitation laser system. 

\section{Surface Effects}

\begin{figure}[htb]
    \includegraphics[width=1.00\columnwidth]{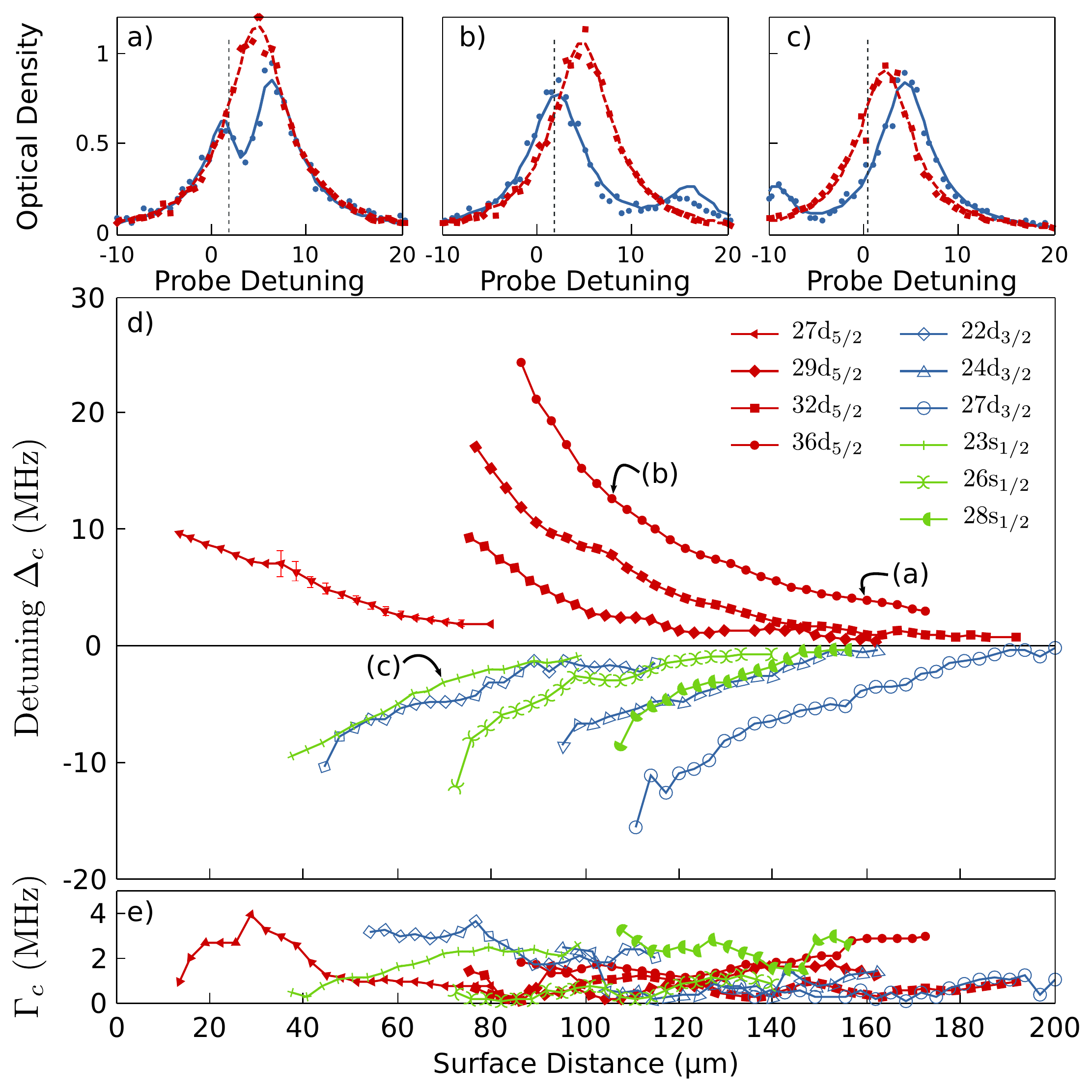}
    \caption{Panels (a)~--~(c) show absorption spectra for three different points marked in the main figure (d). Markers and colors correspond to those used in figure~\ref{fig:spectrum}. The vertical lines in these figures denote the resonance position at large distances. The main figure (d) shows measured shifts of the EIT resonance for various excited states as a function of distance from the surface. The measurement of 27d$_{5/2}$ indicates typical error bars (standard deviation) calculated from repeated measurements of the same state over a time-span of 3 weeks. The lower panel (e) shows fitted values for $\Gamma_c$ for all states presented.} 
    \label{fig:detuning}
\end{figure}

Figure~\ref{fig:detuning} shows the fitted detuning of the EIT resonances as a function of distance to the surface for various excited states in the range $n = 22-36$. Example spectra with fits for some observed level shifts are shown in figure~\ref{fig:detuning}(a,b) for the state 36d$_{5/2}$ and in figure~\ref{fig:detuning}(c) for the state 23s$_{1/2}$. The dashed vertical line indicates the extrapolated position of the EIT resonance for large distances (subtracted in Fig.~\ref{fig:detuning}d); a significant distance-dependent shift of the fitted resonance position is clearly recognizable. A positive shift of the detuning is equivalent to a shift of the Rydberg level towards higher energies.

We investigate s-states $\left|\ell=0, j=\frac{1}{2}\right>$ as well as d-states where $\left|\ell=2, j=\frac{5}{2}\right>$ and $\left|\ell=2, j=\frac{3}{2}\right>$. The distance range probed in the experiments is between \unit{20}{\micro\meter} and \unit{200}{\micro\meter} and in this range we observe shifts between \unit{-15}{\mega\hertz} and \unit{+20}{\mega\hertz}. The measured $n$ dependence is $\propto n^{6.2\pm0.3}$, however there are significant differences between measured shifts for $s$, d$_{3/2}$ and d$_{5/2}$ states.

Notably, the measured shifts for the d$_{5/2}$ states have the opposite sign as compared with the s$_{1/2}$ and d$_{3/2}$ states. This we attribute to the  differing dc electric polarizability $\alpha$ for the studied states indicating the shifts originate from spatially varying electric fields. We compute $\alpha$ from perturbation theory based on the analytical radial matrix elements in terms of Appell F$_2$ functions, including their dependence on $j$ and $|m_j|$ quantum numbers~\cite{Kostelecky1985}, using precise values for the $^{87}$Rb quantum defects~\cite{Li2003}. The calculated values are in excellent agreement with tabulated measurements for the scalar and tensor polarizabilities from previous Stark shift measurements of $s$ and $d$-states of $^{85}$Rb~\cite{O'Sullivan1985, O'Sullivan1986}. In particular we find that for nd$_{5/2, |m_j|=1/2}$, $\alpha$ is negative for the measured states, whereas the $s$ and d$_{3/2, |m_j|=3/2}$ states have a positive sign. From the measured shifts we conclude that the dominant states probed in these experiments are s$_{1/2,1/2}$, d$_{3/2, 3/2}$ and d$_{5/2, 1/2}$. The other $|m_j|$ states are expected to shift at a different range of distances than those investigated, and could have an influence for larger distances. For example, for the state 22d$_{5/2}$ the electric polarizability $\alpha$ for $|m_j|=\frac{3}{2}$ and $|m_j|=\frac{5}{2}$ is about 3 and 10 times larger respectively than for the probed state $|m_j|=\frac{1}{2}$.

The spectra also provide information on the Rydberg lifetime through the parameter $\Gamma_c$. As can be seen in figure~\ref{fig:detuning}(e) the fitted values of $\Gamma_c$ do not measurably depend on the distance to the chip surface. The observed linewidths are of the order of the linewidth measured at large distances to the chip, which is limited by the linewidth of our laser system (see figure \ref{fig:spectrum}a). For some of the data we see evidence for other $|m_j|$ states as double EIT resonances, which may result in artificially large fitted values for $\Gamma_c$. Some broadening of the EIT resonances ($\Gamma_c\sim0.2\Delta_c$) is also expected due to integration over the inhomogeneous electric field over the atomic distribution along the probe axis during imaging. The largest fitted value for $\Gamma_c$ we observed in any measurement was below \unit{4}{\mega\hertz}, giving a lower bound for the Rydberg lifetime of \unit{0.04}{\micro\second}.

\begin{figure}[htb]
	\includegraphics[width=1.00\columnwidth]{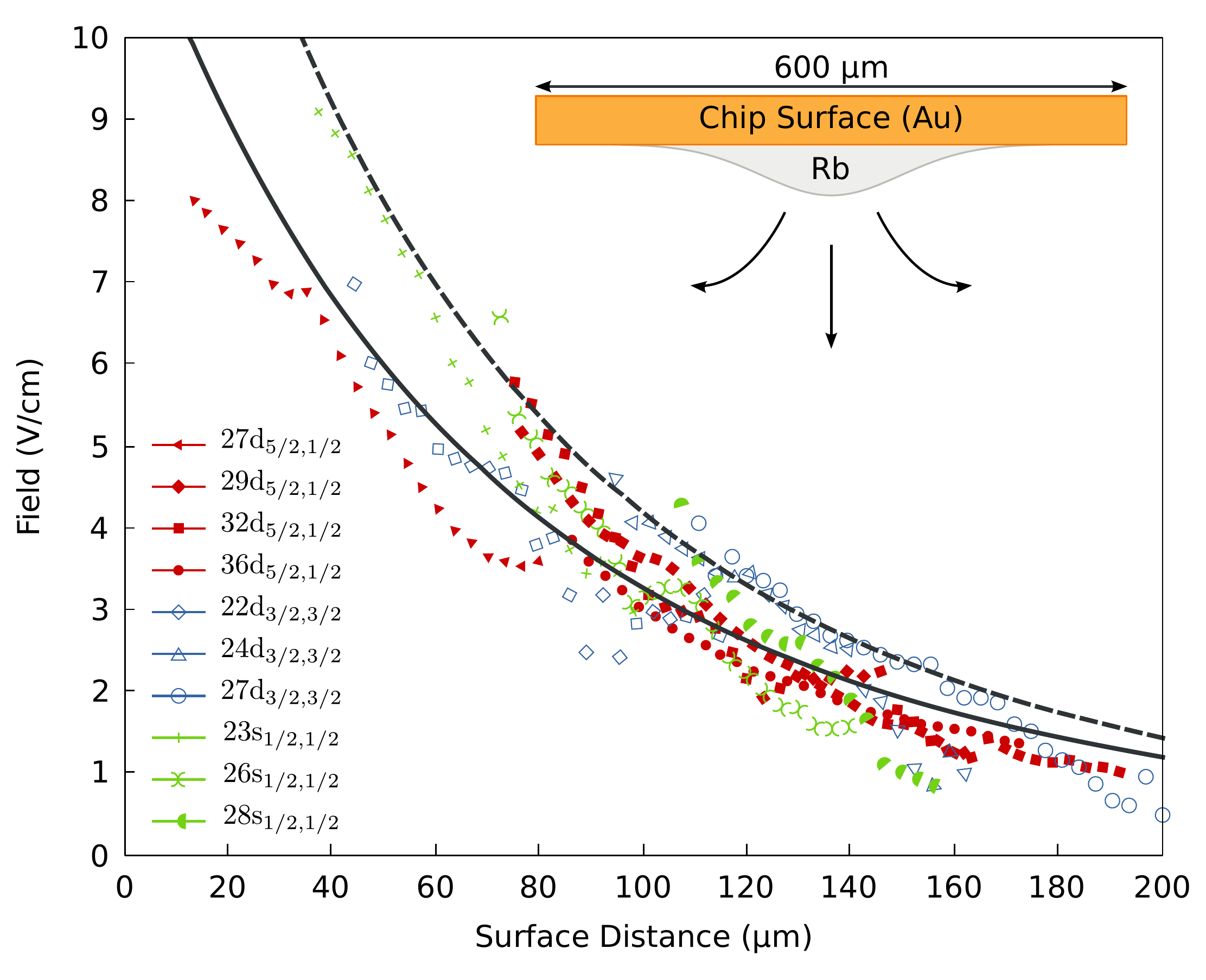}
	\caption{Local electric field strength inferred from the measured Rydberg energy-level shifts and averaged over the probe direction. The solid line shows the results of a fit to the data assuming the field originates from an gaussian patch of surface dipoles (see inset) with $d_0=$ $\unit{7\times10^5}{Debye\per\micro\meter\squared}$, and radius $w=$~\unit{100}{\micro\meter}, averaged over the  $e^{-1/2}$ cloud radius of $\sigma_y=$~\unit{130}{\micro\meter}. The dotted line indicates the peak field strength at the center of the atom cloud ($y=0$).}
	\label{fig:field}
\end{figure}

Shown in Figure~\ref{fig:field} are values for the local electric field strength $|\overline{E}|=\sqrt{2 \Delta_c/\alpha}$ inferred from the measured Rydberg energy-level shifts. The measurements are identical to those presented in figure \ref{fig:detuning}; they span from $n=22-36$ and include $s$, d$_{3/2}$ and d$_{5/2}$ angular momentum states covering a wide range of electric polarizabilities. In this form, all data fall on a single curve, confirming that the measured line shifts are caused by an inhomogeneous electric field originating from the chip surface. The height dependence of the field strength roughly follows a weak power-law decay $\propto z^{-0.7}$, thus excluding effects due to Van der Waals interactions ($\propto z^{-3}$)~\cite{Sandoghdar1992} or randomly distributed patch potentials expected to scale with $z^{-2}$. We also expect a contribution of the Van der Waals interaction for e.g. 26s of only $\sim$~\unit{0.1}{\kilo\hertz} at \unit{80}{\micro\meter}. The presented data was obtained over a time-span of approximately 4 weeks of experiments, indicating the observed electric field is relatively stable and reproducible from day-to-day.

We account for our measurements by assuming the electric field is produced by a patch of Rb adsorbates deposited on the chip surface. Following~\cite{Obrecht2007}, we treat each adsorbate as an electric dipole oriented perpendicular to the chip surface and assume a patch of adsorbates which reflects the gaussian distribution of atoms released from the magnetic trap. 
Integrating the field over the distribution of dipoles gives the total electric field as a function of distance from the surface. At the center of the patch the field is given by $E_z(Z)\!=\!\frac{d_0}{2 w \epsilon_0}\!\times\!\left[-Z+e^{\frac{1}{2}Z^2}\sqrt{\frac{\pi}{2}}\left(1+Z^2\right) \text{Erfc}\left(\frac{Z}{\sqrt{2}}\right)\right]$ where $Z=z/w$, $d_0$ is the peak dipole density and $w$ is the $e^{-1/2}$ patch radius. Finally, we include averaging of the measured line shifts over the width of the atom cloud $\sigma_y=$~\unit{130}{\micro\meter} (along the probe axis at the time of detection). This model is then fit to the data by least squares minimization as a function of $z$ to extract the dipole density and patch size. The solid-line in Fig.~\ref{fig:field} shows the result of a fit to the entire data set. The fit parameters are a peak dipole density $d_0=$~\unit{7\times 10^5}{Debye\per\micro\meter\squared} and a patch size $w=$~\unit{100}{\micro\meter}, which is comparable to the radius of the atom cloud after expansion and the distance at which the trapped cloud is released. The dashed-line indicates the inferred field strength at the center of the cloud which is roughly 20\% larger than the cloud averaged field strength. 

From the fit results and for a dipole moment of order \unit{10}{Debye\per adatom}~\cite{Obrecht2007} we estimate a total of $\sim10^9$ adsorbates on the surface. Assuming at steady state a decay rate of the order of $\unit{2\times 10^{-6}}{\reciprocal\second}$ (room temperature surface~\cite{Obrecht2007}) and an experimental cycle time of \unit{30}{\second}, this corresponds to $3\times 10^5$ deposited atoms per shot, comparable to the number of atoms released in each run of the experiment ($N=7\times 10^5$). Since these measurements, we have attempted to clean the surface of adsorbates by resistively heating the Z-shaped wire beneath the chip surface ($I_z=$~\unit{14}{\ampere}, conductor temperature $T\sim$~\unit{80}{\celsius}) for 48 hours, however this was insufficient to significantly raise the surface temperature and did not have an appreciable effect on the measured Rydberg lineshifts. Briefly exposing the surface to oxygen may be a better method to elimate the electric field produced by rubidium adsorbates.

\section{Conclusion}
The present experiments focus on spatially resolved excitation of Rydberg atoms on an atom chip. We create Rydberg atoms in a cloud of rubidium atoms on an atom chip, and investigate electromagnetically induced transparency near the chip surface. Absorption imaging in conjunction with the recording of EIT signals is used to obtain spatially resolved EIT spectra. We measure significant shifts of the Rydberg levels as we approach the surface which we attribute to electric fields produced by adatoms adsorbed on the chip surface. A theoretical model of the field produced by a gaussian distribution of surface dipoles is in good agreement with our data. We do not observe significant broadening of the Rydberg levels. The measured shifts are not expected to inhibit the coherent creation and investigation of Rydberg atoms on atom chips.

We clearly see that the shifts are directly proportional to the polarizabilities of the states in question. This opens new possibilities to use Rydberg excited atoms as a sensitive, spatially resolved probe of electric fields~\cite{Osterwalder1999, Kocher1987}. A simple estimate of our present sensitivity to electric fields is $\sim$~\unit{0.1}{\volt\per\centi\meter} with a resolution \unit{7}{\micro\meter}. The field sensitivity could be straight forwardly improved using narrow linewidth lasers and higher-lying Rydberg states. 

The surface effects due to adatoms observed here could be prevented in future experiments by incorporating a magnetic field gradient to push the atoms away from the surface at the end of each experimental cycle. Furthermore, use of other coating materials on the chip surface could decrease the dipole moment of adatoms, or increase the desorption rate of these atoms.

Highly excited Rydberg atoms on atom chips, as demonstrated here for the first time, open new avenues for the study of dipolar physics, strongly interacting systems, and quantum information science in tailored trapping potentials. Atom chips allow the preparation of small atomic clouds~\cite{Whitlock2009} ideal for investigating dipole-dipole interactions and collective excitations in dense (\unit{\power{10}{15}}{\centi\meter\rpcubed}) mesoscopic ensembles~\cite{Saffman2009,Muller2009a}. Furthermore, Rydberg atoms on atom chips are very attractive for studying long-range Van der Waals interactions with surfaces~\cite{Sandoghdar1992} and interfacing ultracold atoms with on-chip structures~\cite{Sorensen2004, Andre2006}.

\begin{acknowledgments}
We thank N. J. van Druten and J-M. Raimond for valuable discussions. This work is part of the research 
programme of the 'Stichting voor Fundamenteel Onderzoek der Materie (FOM)', 
which is financially supported by the 'Nederlandse Organisatie voor 
Wetenschappelijk Onderzoek (NWO)'. SW acknowledges support from a Marie-Curie fellowship (PIIF-GA-2008-220794).
\end{acknowledgments}

\appendix

\end{document}